\documentclass[preprint,12pt,3p]{elsarticle}

\usepackage{amssymb}
\usepackage{amsmath}

\usepackage[justification=centering]{caption}
\usepackage{subfig}
\usepackage{bm}
\usepackage[ruled]{algorithm2e}
\usepackage{pgfplots}
\pgfplotsset{compat=1.15}
\usepgfplotslibrary{fillbetween}
\usetikzlibrary{external}
\tikzset{external/system call={pdflatex \tikzexternalcheckshellescape -halt-on-error -interaction=batchmode --extra-mem-bot=999999999 --extra-mem-top=999999999 -jobname "\image" "\texsource"}}
\usetikzlibrary{decorations.pathmorphing,mindmap,backgrounds,arrows.meta,positioning,shapes,calc}

\journal{Mechanical Systems and Signal Processing}

\AtBeginDocument{%
  \let\mathbb\relax
  \DeclareMathAlphabet\PazoBB{U}{fplmbb}{m}{n}
  \newcommand{\mathbb}{\PazoBB}
}

\begin{document}
\newcommand{\change}[1]{#1}
\begin{frontmatter}

\title{Intelligent Optimization and Machine Learning Algorithms for Structural Anomaly Detection using Seismic Signals}

\author[label1]{Maximilian Trapp\corref{cor1}}
\ead{Maximilian.Trapp@rub.de}

\author[label2]{Can Bogoclu}
\ead{Can.Bogoclu@hs-niederrhein.de}

\author[label1]{Tamara Nestorovi\'{c}}
\ead{Tamara.Nestorovic@rub.de}

\author[label2]{Dirk Roos}
\ead{Dirk.Roos@hs-niederrhein.de}

\address[label1]{AG Mechanics of Adaptive Systems, Ruhr-University Bochum, Building ICFW 03-725, Universit\"atsstra\ss{}e~150, 44801~Bochum, Germany}
\address[label2]{Institute of Modeling and High-Performance Computing, Niederrhein University, Reinarzstra\ss{}e~49, 47797~Krefeld, Germany }

\cortext[cor1]{I am the corresponding author. Interested readers may contact me regarding structural anomaly detection and unscented hybrid simulated annealing and C. Bogoclu regarding machine learning.}

\begin{abstract}
The lack of anomaly detection methods during mechanized tunnelling can cause financial loss and deficits in drilling time. 
On-site excavation requires hard obstacles to be recognized prior to drilling in order to avoid damaging the tunnel boring machine and to adjust the propagation velocity. The efficiency of the structural anomaly detection can be increased with intelligent optimization techniques and machine learning. In this research, the anomaly in a simple structure is detected by comparing the experimental measurements of the structural vibrations with numerical simulations using parameter estimation methods. 
\end{abstract}


\end{frontmatter}



\section{Introduction}
\label{sec1}
Drilling into unknown soil bears a large economical risk. Possible unfavourable scenarios span from excess water inflow or a damaging of the Tunnel Boring Machine (TBM) to a total collapse of the tunnel \cite{Schmitt2004}. To avoid potential risks, the imaging of voids, faults, fluid areas, erratic boulders or other changes in material is essential. Exploratory drillings only provide an image of the geological parameters in the near-field of the borehole and lack in showing the detailed geological structure. Thus, acoustic analysis is a better choice as it offers the opportunity to obtain a detailed image of the soil with the help of seismic waves.~Propagating through the ground, seismic waves are reflected, refracted, scattered and converted; resulting in a detailed fingerprint of the actual structure. 

Most of the techniques used nowadays for the detection of anomalies rely on travel time measurements and migration techniques, considering only compressional waves. Therefore, typical state-of-the-art systems like Tunnel Seismic Prediction (TSP) \cite{Sattel1992}, Sonic Softground Probing (SSP) \cite{Kneib_etal2000}, True Reflection Tomography (TRT) \cite{Otto_etal2002} or Integrated Seismic Imaging System (ISIS) \cite{Borm_etal2003} lack accuracy in describing a detailed image of the subsoil. Aiming at full exploitation of the information coming from seismic waves, full waveform inversion (FWI) techniques \cite{Tarantola1984} offer promising opportunities for reconnaissance in mechanized tunneling. Musayev et al. \cite{Musayev2014} propose an algorithm for forward modeling and inversion of seismic waves in frequency domain for 2D and 3D tunnel models. Bretaudeau et al.~use an iterative conjugate-gradient approach in frequency domain for 2D models, which was already tested on experimental data \cite{Bretaudeau_etal2013, Bretaudeau_etal2014}. Also, the structural composition of rocks was investigated on the microscale in relation with detection and prediction of failure and deterioration  \cite{du2017microstructure}. A nodal discontinuous Galerkin method implemented into an adjoint approach is used by  Lamert \& Friederich \cite{Lamert2018} to achieve FWI in time domain for 2D models.

Nguyen \& Nestorovic \cite{Nguyen2016} propose unscented hybrid simulated annealing (UHSA) for FWI in time domain. Here, the unscented Kalman filter (UKF) \cite{Julier&Uhlmann2004} and simulated annealing (SA) \cite{Kirkpatrick_etal1983} are combined to achieve a fast minimization of the misfit function. As the inversion parameters are free to choose in this algorithm, a large portion of the parameter space can be reduced if the shape of the structural anomaly is roughly known. Another advantage of UHSA compared to gradient-based methods is that the starting model does not have to be close to the real model as the algorithm bases on the non-deterministic exploration of all parameter configurations. Hence, UHSA conducts a global search and is not prone to the effects of local minima like the gradient-based methods. 

Supervised machine learning (SML) methods are deployed to decrease the number of required calculations for the solution of the structural anomaly detection problems. This is achieved through the replacement of the computationally burdensome numerical model with a statistical model, based on the calculations conducted on the former. The prediction time of a SML model is expected to be smaller than a second, depending on the used method, the amount of data and the number of dimensions. This allows deploying greedy optimization strategies to acquire the global optimum. In the machine learning literature, the term anomaly detection is often used in terms of detecting the anomalies in the data. However, the aim of this research is the detection of structural anomalies using a numerical model, hence the task is essentially a parameter estimation problem. A random search algorithm combined with artificial neural networks is proposed in \cite{Brigham2007} for inverse material identification. Kriging is used not to replace the numerical model but to directly approximate an inverse model for electromagnetic non-destructive testing in \cite{Bilicz2012}. In \cite{Marzouk2009}, Gaussian process is combined with polynomial chaos expansion for the solution of the inverse problems using Bayesian inference. In \cite{Aster2013}, the use of nonlinear regression is proposed for inverse problems. All of these methods as well as various other approaches based on machine learning can be used for the structural anomaly detection problems, but most of these methods approximate single-output or scalar responses. Hence for the evaluation of signals, the models are used to approximate some integral error metric and not the discrete temporal signals.

However, using multi-output machine learning methods to approximate time-dependent responses is meaningful for the structural anomaly detection using acoustic signals. Parameter estimation of signals using estimate maximize algorithm is proposed in \cite{Feder1988}, where the signals are decomposed and each component is analysed separately.~In \cite{Santamarina2005}, various approaches for the solution of inverse problems regarding signals are discussed, where the possibility of using artificial neural networks is briefly introduced.~Nevertheless, there is considerably less research regarding the use of the multi-output supervised machine learning for inverse problems with time-dependent data.

The structural anomaly detection potential using FWI and SML methods is investigated on a surrogate laboratory model, which provides real measurement data.~Thus, the laboratory offers an environment for the investigation of signal features that are in the sense of data acquisition close to real conditions. In contrast, pure numerical approaches often neglect statistical issues or require artificial signal contamination by some known pseudorandom noise. In this research, the eligible data for FWI is generated from experiments conducted on an aluminium block with a drilling hole. Thereupon, a spectral element model of the structure with a sufficient prognosis quality of the waveforms is prepared for the FWI. The determination of the model parameters as well as the estimation of the source signature of the transducer is decisive for a successful inversion. Finally, UHSA and deep Gaussian covariance network (DGCN) are compared against a generic black box optimization algorithm for the determination of the hole coordinates. 

The described procedure is of high importance for validation purposes of the constructed small-scale experiment on the one hand as well as for the developed algorithms on the other hand. Here, aluminium can be regarded as an elastic, homogeneous and isotropic medium. For real soil materials, which shall be investigated in future research, an inversion can be more challenging, as heterogeneities, anelasticity and anisotropy are some of the difficulties one needs to encounter. Including these properties into the model, the inversion parameter
space gets higher, meaning that more computational resources are required.

\section{Underlying Methods}
\label{sec2}
A general representation of the nonlinear quasi-static numerical model can be given as follows:
\begin{equation}
	\bm{d} = h\left( \bm{m}\right) + \bm{v},
\end{equation}
where  $\bm{m} \in \mathbb{R}^n$ is the state vector containing $n$ parameters which shall be used to describe the anomaly in the model. The vector $\bm{d} \in \mathbb{R}^r$ stores the model outputs of the current time step for $r$ receivers arising from the evaluation of the nonlinear simulation function $h: \mathbb{R}^n \mapsto \mathbb{R}^r$.  $\bm{v} \in \mathbb{R}^r$ is the modeling error resulting from assumptions made in describing the real model as well as the error resulting from numerical approximations. The misfit $s_{j}: \mathbb{R}^n \mapsto \mathbb{R}$ at the receiver point $j$ is defined by summing the squared differences between the measurement  $d^{\mathrm{obs}}_j(t)$ and the model output $d_j(\bm{m},t)$ of the simulation in the time window $[0, \tau]$ as follows:
\begin{equation}
\label{eqn:misfit}
	s_{j} = \frac{1}{2} \int \limits_0^{\tau} \| d^{\mathrm{obs}}_j(t) - d_j(\bm{m},t) \|^2 dt  ,j \in \{1,2,..,r\}.
\end{equation}
In this section, two methods are presented that aim at minimizing these misfits by finding the optimal configuration of the model parameters contained in vector $\bm{m}$ efficiently. Generally, the optimization problem can be solved using generic optimization algorithms such as the greedy particle swarm optimization (PSO) to find the global optimum.~However, this is generally not applicable to on-site problems since the computation time is an important constraint. To tackle this problem, the novel unscented hybrid simulated annealing algorithm is described, which is a non-deterministic approach that combines a global optimization approach with a local minimization technique. This increases the efficiency of the optimization. Furthermore, DGCN is introduced as a supervised machine learning method, which reduces the runtime of the simulation by replacing the simulation model using a finite amount of calculated samples.

\subsection{Unscented Hybrid Simulated Annealing}
\label{subsec21}

\subsubsection{Concept}

\begin{figure} [ht]
\center 
\includegraphics[width=0.5\textwidth]{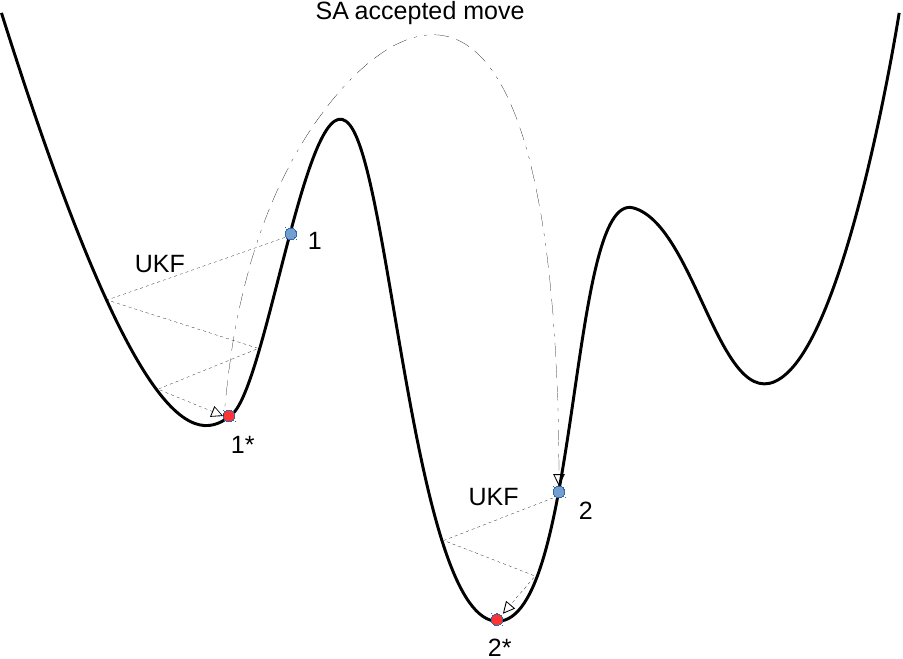}
\caption{Principle of the UHSA algorithm \cite[p.~238]{Nguyen2016} }
\label{fig:UHSA_concept}
\end{figure}

Section \ref{subsec21} describes the working principle of the algorithm of unscented hybrid simulated annealing (UHSA) proposed by Nguyen \& Nestorovic \cite{Nguyen2016}. The global optimization approach of simulated annealing (SA) developed by Kirkpatrick et al.~\cite{Kirkpatrick_etal1983} is combined with the unscented Kalman filter (UKF) \cite{Julier&Uhlmann2004} for local minimization. The developed algorithm aims at the quick exploration of the full parameter space of the underlying inverse problem. The overall concept of the UHSA algorithm is illustrated on a multimodal misfit landscape in a one-dimensional space (Figure \ref{fig:UHSA_concept} \cite[p.~238]{Nguyen2016}).
Beginning with a random sample (1), the UKF exploits the region around the sample within a user-defined number of steps. The higher the number of steps is set, the more precise is the estimation of the local minimum. When the steps are completed and the misfit of the parameter configuration is close to the local minimum (1*), a new random step is proposed (2). If the misfit of this step is better than the best misfit found up to that point, the move is accepted. If not, the move is only accepted with a certain probability. If the proposed sample is close to the global minimum like in (2), the SA algorithm accepts it because of the low misfit. The UKF investigates this region and the global minimum will be found with a precision that is dependent on the user-defined number of Kalman filter steps.

\label{subsec211}
\subsubsection{Unscented Kalman Filter for Local Minimization}
\label{subsec212}

The UKF is a derivative-free estimator of nonlinear state-space models. Although the extended Kalman filter (EKF) is the most widely used algorithm for the state estimation of nonlinear systems, it may fail at finding the global optimum if the nonlinearity is high \cite{Simon2006}.~This is caused by the linearization around the current mean, which is implemented in the EKF and which produces large errors in the posterior estimates if the investigated model area is not close to linear. The UKF does not need a linearization as it models the nonlinearity by approximating the distribution of the state random variable \cite{VanDerMerwe2004}.~More specifically, it is based on the propagation of the mean and covariance of the estimated quantities by a set of sampled points which are representing the state distribution, the so-called sigma-points.~A set of $2n+1$ sigma-point vectors $\bm{\mathcal{M}}_i$ is computed as follows:
\begin{eqnarray}
\label{eqn:sigma_points}
			\bm{\mathcal{M}}_0 &=& \hat{\bm{m}},  \\
			\bm{\mathcal{M}}_i &=& \hat{\bm{m}} + \left(\sqrt{(n+\lambda)\bm{P}^m}\right)_i \textrm{, for } i = 1:n,  \\
			\bm{\mathcal{M}}_{n+i} &=& \hat{\bm{m}} - \left(\sqrt{(n+\lambda)\bm{P}^m}\right)_i \textrm{, for } i = 1:n,  
\end{eqnarray}
where $\hat{\bm{m}}$ is the mean of the random variable with $n$ model parameters and the covariance $\bm{P}^m$. The parameter $\lambda$ is a scaling parameter which is free to tune but is suggested to select $n+\lambda=3$ if $\bm{m}$ is Gaussian \cite{Julier&Uhlmann2004}. $(\cdot)_i$ is the $i$-column of the matrix. In order to calculate the square root of $\bm{P}^m$, Cholesky factorization is used. \\
Misfit vectors $\bm{\mathcal{S}_i}$ are calculated by evaluation of Eq. \ref{eqn:misfit} at each sigma-point:
\begin{equation}
\label{eqn:nonlinear_mapping}
	\bm{\mathcal{S}}_i = \bm{s}\left(\mathcal{M}_i\right) \textrm{, for } i = 0:2n.
\end{equation}
Evaluation produces a transformation $\bm{s}: \mathbb{R}^n \mapsto \mathbb{R}^r$ for $r$ receivers with $\bm{s} = (s_1, s_2,\dots,s_r)$. Each sigma-point is associated with a weight $W_i$ so that the sum of the weights is unity:
\begin{eqnarray}
	W_0 &=& \frac{\lambda}{n+\lambda}, W_i = W_{i+n} = \frac{1}{2(n+\lambda)} \textrm{, for } i = 1:n.
	\label{eqn:weights}
\end{eqnarray}
The approximated mean of the misfit is obtained by a weighted summation of the misfit vectors:
\begin{eqnarray}
	\hat{\bm{s}} &=&  \displaystyle\sum\limits_{i=0}^{2n} W_i  \bm{\mathcal{S}}_i. \label{eqn:mean_d}
\end{eqnarray}
In order to obtain a parameter prediction, we update the covariance of the parameters $\bm{P}^m$, the covariance of the misfit $\bm{P}^s$ and the cross-covariance $\bm{P}^{ms}$ between them by a weighted summation as follows:
\begin{eqnarray}
	\bm{P}^m &=&  \displaystyle\sum\limits_{i=0}^{2n} W_i  \left(\bm{\mathcal{M}}_i - \hat{\bm{m}}\right) \left(\bm{\mathcal{M}}_i - \hat{\bm{m}} \right)^t + \bm{Q}, \label{eqn:covariance_Pm}\\
	\bm{P}^s &=&  \displaystyle\sum\limits_{i=0}^{2n} W_i  \left(\bm{\mathcal{S}}_i - \hat{\bm{s}}\right) \left(\bm{\mathcal{S}}_i - \hat{\bm{s}} \right)^t + \bm{R}, \label{eqn:covariance_Ps}\\
	\bm{P}^{ms} &=&  \displaystyle\sum\limits_{i=0}^{2n} W_i  \left(\bm{\mathcal{M}}_i - \hat{\bm{m}}\right) \left(\bm{\mathcal{S}}_i - \hat{\bm{s}} \right)^t. \label{eqn:covariance_Pms}
\end{eqnarray}
Process noise covariance matrix $\bm{Q}$ adds an amount of Gaussian noise to the covariance of the parameters. By setting the values of this matrix, the user can prevent distances between the sigma-points from getting too small during the optimization process. Every set of measurement is contaminated with noise which can be described by a measurement covariance matrix $\bm{R}^{\mathrm{meas}}$. Additionally, we have a modeling uncertainty which can be expressed by a covariance matrix $\bm{R}^{\mathrm{model}}$. Thus, covariance matrix $\bm{R}$ can be described as $\bm{R} = \bm{R}^{\mathrm{meas}} + \bm{R}^{\mathrm{model}}$.\\
Posterior mean of the sought parameters $\hat{\bm{m}}_{+}$ and its covariance $\bm{P}^m_{+}$ are updated as follows:
\begin{eqnarray}
  	\hat{\bm{m}}_{+} &=& \hat{\bm{m}} + \bm{G} \left( \bm{s}_{min} - \hat{\bm{s}} \right), \label{eqn:correction_m}\\
	 \bm{P}^m_{+} &=& \bm{P}^m - \bm{G}  \bm{P^}s \bm{G}^t, \label{eqn:correction_Pm}						
\end{eqnarray}
with the Kalman gain $\bm{G}$ calculated as
\begin{equation}
	\bm{G} = \bm{P}^{ms} \left(\bm{P}^s \right)^{-1}. \label{eqn:Kalman_gain_UKF} 
\end{equation}   	

The term $\bm{s}_{min} - \hat{\bm{s}}$ is called the innovation term, where $\bm{s}_{min}$ is the best expected result. To find the local minimum with a high precision, the UKF is run multiple times. The tuning of this number is application-specific as it is dependent on the calculation time and the required precision (see Section \ref{subsec51} for further discussion). In the next section, the UKF is integrated into the algorithm of SA to achieve global optimization.

\subsubsection{Unscented Hybrid Simulated Annealing}
\label{subsec213}

SA is a probabilistic and global optimization technique. With their Metropolis algorithm, Metropolis et al.~\cite{Metropolis_etal1953} form the base of SA. The complete algorithm is later introduced into statistical mechanics by Kirkpatrick et al.~\cite{Kirkpatrick_etal1983}, which thus provides an alternative optimization possibility for complex systems.

Staying trapped in local minima is a common problem for multi-dimensional and highly nonlinear spaces. The Metropolis algorithm provides a solution for escaping from these minima in order to find the global solution. The key of the concept is that a new sample can be accepted although the solution of this is less favourable than the current result. Acceptance probability depends on how bad the solution is compared to the current result as well as on the number of iterations so far, expressed by the so-called annealing temperature. If a solution is better than the current solution, the move is always accepted.

At the beginning of every annealing cycle, a random configuration of parameters is proposed by 
\begin{equation}
\label{eqn:Annealing_generator}
	m_c^i = L^i+  u^i\left(U^i - L^i\right),
\end{equation}
where $u^i$ is drawn from a uniform distribution $u^i \in [0,1]$ and $U^i$ and $L^i$ are the upper and lower bound of the current parameter configuration $m_c^i$.
For this parameter configuration, a misfit functional is calculated by summation of the misfit values at each receiver:
\begin{equation}
\label{eqn:eqn:misfit_functional}
	S = \displaystyle\sum\limits_{j=1}^{r} s_j.
\end{equation}
The probability of acceptance $\mathcal{A}$ is calculated as proposed in \cite{Balling1991}:
\begin{equation}
\label{eqn:acceptance_prob}
	P(\mathcal{A}) = \textrm{exp}\left(-\frac{\Delta S}{\Delta \overline{S} \, T_c}\right),
\end{equation}
where $T_c$ is the current annealing temperature. In our case, $\Delta S$ is chosen to be the misfit difference between the proposed parameter set and the best result up to that point.  $\Delta \overline{S}$ is the average of all $\Delta S$ that are accepted.  If the current parameter set is accepted by the SA, the UKF exploits the region. As proposed by Blum \& Roli \cite{Blum&Roli2003}, the cooling schedule is selected to be linear as:
 \begin{equation}
\label{eqn:cooling}
	T_{c+1} = \alpha T_c,
\end{equation}
where $\alpha \in (0,1)$. The resulting algorithm is slightly different than in Nguyen \& Nestorovic \cite{Nguyen2016} and is summarized in Algorithm 1.

\begin{algorithm}[H]
	\DontPrintSemicolon
 	\tcp{Initialization}
 	Settings for the UKF: $\bm{P}_0, \bm{R}, \bm{Q}, N_k$\;
 	Settings for the SA: $T_0, \alpha, N_c$\;
  $c \gets 1$\;  
  \tcp{run $N_c$ cycles of SA}
	\While{$c < N_c$} {
		Select random configuration by Eq.~\ref{eqn:Annealing_generator}\;	
		Calculate acceptance probability $P$: Eq.~\ref{eqn:acceptance_prob} \\
		Check if the proposed move is accepted:\\
		\If{$\textrm{random}.\textrm{uniform}(0,1) < P$}{
			\tcp{Run the UKF for $N_k$ iterations}
			\For{$k = 1..N_k$} {
				 $\hat{\bm{m}}_k, \bm{P}_k = \textrm{UKF}\left(\hat{\bm{m}}_{k-1}, \bm{P}_{k-1},\bm{R},\bm{Q} \right)$\;				
		 	}
		 	Store the improved point $\bm{m}_c = \hat{\bm{m}}_k$\;
		}		
  	Lowering temperature $T_c$: Eq.~\ref{eqn:cooling}\;  	
  	$c \gets c + 1$\;
	}
	\caption{unscented hybrid simulated annealing}
	\label{alg:UHSA}
\end{algorithm}

\subsection{Supervised Machine Learning}
\newlength{\fheight}%
\newlength{\fwidth}%

Machine learning algorithms are computer programs that can \textit{learn} from the experience without being explicitly reprogrammed. Formally, a computer program is said to learn from experience (i.e. the data) $E$ with respect to some class of tasks $T$ (such as regression, classification) and the performance measure $P$ (such as approximation quality, classification success rate, cross entropy), if its performance at tasks $T$, as measured by $P$ improves with experience $E$ \cite{Mitchell1997}.

There are at least two different categories for machine learning, that describe a dichotomy between supervised and unsupervised learning as well as online and offline learning. The first category describes, if the data is labelled, meaning each input point in the training set of points corresponds to an output. Algorithms that learn from labelled data are regarded as supervised, whereas the unsupervised algorithms seek to find dependencies in the unlabelled data. The second category expresses the capability of the model to be updated after the initial training is finished, also known as the online learning. Supervised machine learning algorithms in this research are trained offline to build a surrogate model of the computationally expensive numerical model.

\subsubsection{Deep Gaussian Covariance Network}
The choice of the best supervised machine learning (SML) algorithm is not a trivial one. Many of the contemporary methods can deliver a good approximation, depending on the type of the problem but it is not an easy task to find a method that outperforms others consistently. A benchmark study of some supervised learning methods can be found in \cite{Bogoclu2016}. Anisotropic Gaussian process (GP) \cite{Rasmussen2006} is one of the most popular methods for supervised machine learning. The prediction function $\tilde{f}(\mathbf{x})$ of a GP for the $n$-dimensional input design of experiments (DoE) $\mathbf{X}^0$ with $N$ samples and the response vector $\mathbf{y}^0$
\begin{equation}
\begin{gathered}
\tilde{f}(\mathbf{x}) = \mathbf{P}_R(\mathbf{x},d) \hat{\mathbf{c}}  + R(\mathbf{x}) \\ 
R(\mathbf{x}) =\mathbf{k}(\mathbf{x}) \tilde{\mathbf{K}}(\mathbf{X}^0)^{-1}(\mathbf{y}^0 - \mathbf{P}_R(\mathbf{X}^0,d)^T \hat{\mathbf{c}}) 
\end{gathered}
\end{equation}
consists of two parts. The first part is a polynomial regressor for the global estimation of the mean value at the prediction point $\mathbf{x}$. The order of the polynomial $\mathbf{P}_R(\mathbf{x},d) \hat{\mathbf{c}}$ is often chosen as 0, since the variability of the approximated function can often be described better with the chosen covariance kernel $k_i(\mathbf{x})$ such as the squared exponential function. 
\begin{equation}
k_i(\mathbf{x},\boldsymbol{\theta}) = e^{\left( -\sum\limits_{j=1}^n \theta_j \left( x_{j} - x^0_{i,j} \right)^2 \right)}
\end{equation}

There are various covariance kernels and the choice of the best kernel function has a large impact on the quality of the resulting model, along with obtaining the optimal tuning parameter $\boldsymbol{\theta}$. 

\begin{figure}[ht]%
\setlength\fheight{0.3\textwidth}%
\setlength\fwidth{0.7\textwidth}%
\centering

\includegraphics{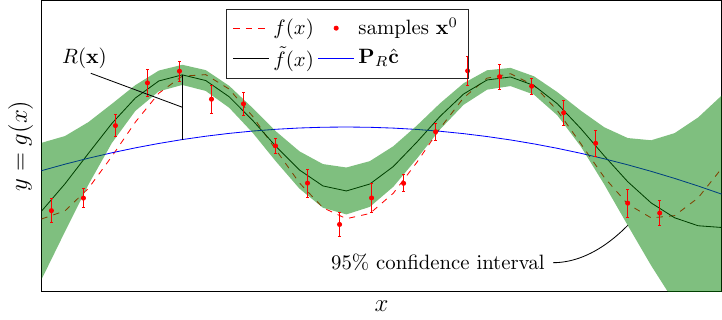}
\caption{Gaussian process approximation with noisy data}
\label{fig:GPdemo}%
\end{figure}

The second part $R(\mathbf{x})$ interpolates the errors between the approximation of the polynomial term and the data points with the help of the covariance kernel $\mathbf{k}(\mathbf{x})$ as shown in Figure \ref{fig:GPdemo}.  $\mathbf{k}(\mathbf{x}) = [k_1,k_2,\dots,k_N]$  and the autocovariance matrix $\mathbf{K}(\mathbf{X}^0)$ consist of the covariances for the input DoE $\mathbf{X}^0$. 
\begin{equation}
\mathbf{K}(\mathbf{X}^0) = 
\begin{pmatrix}
\begin{array}{rrrr}
k_1(\mathbf{x}^0_1) & k_1(\mathbf{x}^0_2) & \cdots & k_1(\mathbf{x}^0_N) \\
k_2(\mathbf{x}^0_1) & k_2(\mathbf{x}^0_2) & \cdots & k_2(\mathbf{x}^0_N) \\
\vdots & \vdots & \vdots & \vdots \\
k_N(\mathbf{x}^0_1) & k_N(\mathbf{x}^0_2) & \cdots & k_N(\mathbf{x}^0_N)
\end{array}
\end{pmatrix}
\end{equation}

Both of these functions are dependent on the bandwith or length scale parameter $\boldsymbol{\theta}$, which is obtained during the \textit{training}. Furthermore, a noise term $\eta$ is introduced to the covariance matrix to bound the pure interpolation and create a smoother surrogate model as in
\begin{equation}
\tilde{\mathbf{K}}(\mathbf{X}^0) = \mathbf{K}(\mathbf{X}^0) + \eta \mathbf{I}_N
\end{equation}
where $\mathbf{I}_N$ denotes the $N \times N$ identity matrix. This is especially useful for practical examples, if the data has some intrinsic noise because of e.g. measurement errors, variable selection or non-convergent CAD simulations. All tuning parameters of the surrogate model $\eta$ and $\boldsymbol{\theta}$ as well as the choice of the best covariance function $k_i$ can be obtained through training i.e. optimization of the leave-one-out cross validation of the approximation error \cite{Hastie2009} or maximizing the likelihood \cite{Rasmussen2006}.

\begin{figure}[ht]

\centering
\footnotesize

\includegraphics{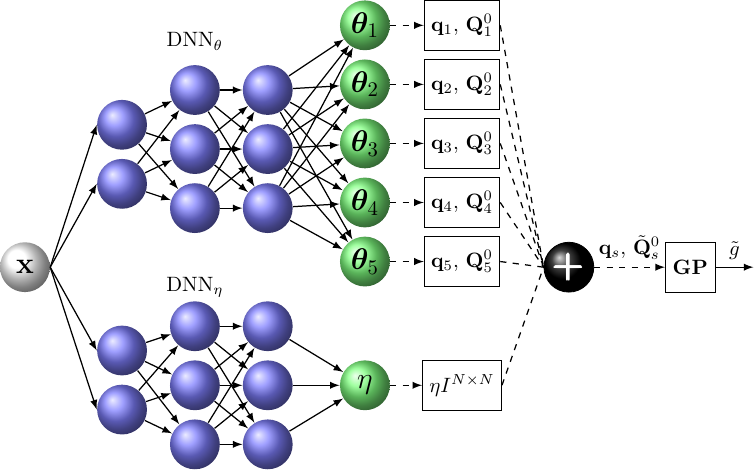}
\caption{Schematic overview of a DGCN. The input dimension is not related to the presented application.}
\label{fig:dgcn_pic}
\end{figure}

The supervised machine learning algorithm used in this research is the DGCN. As the name suggests, DGCN is an artificial neural network that determines the  $\eta$ and $\boldsymbol{\theta}$ for multiple covariance kernels. $\eta$ and $\boldsymbol{\theta}$ in DGCN are not globally constant as in the traditional GP but are predicted by a subsequent neural network at each prediction point (see Figure \ref{fig:dgcn_pic}).  A more detailed description of DGCN is beyond the scope of this research, but \cite{Cremanns2017} is recommended for further reading. 

\subsubsection{Multi-Output Supervised Machine Learning}\label{sec:MOSML}
SML is often used for single-output problems, where $f(\mathbf{x})$ returns a scalar value. However, being able to approximate signals $f(\mathbf{x},t)$ is useful for some applications. In the case of anomaly detection, using single-output SML methods requires the models to be trained regarding a scalar measure of error. The ability of this measure to reflect the characteristic differences between the measured and the simulated signals is important. The measure also affects the quality of the machine learning algorithms, since these try to find a simplified model, that has the same major characteristics as the underlying response. Hence, if the response is not sensitive enough to the characteristic differences in the compared signals, the abstraction may smooth out this behaviour. Furthermore the location of the global optimum depends on the measurement. Each time a new measurement is acquired, the error at each sampling point needs to be recomputed and a new model must be trained. This is challenging for on-site applications. In contrast, if a multi-output SML method is used, the surrogate model approximates the signals directly. Consequently, the same model can be used for any future measurement and the model quality is independent of the used metric for the assessment of the differences.

Several strategies can be adopted to solve multi-output machine learning problems. For example, recurrent neural networks are used to approximate functions in the form $(f(x(t),t)$ where $x(t)$ is constant over all time steps. However, such a strategy requires all of the signals to be composed to a single signal and use of auxiliary variables to separate different signals, which can be exhausting for applications with larger data sets.

\begin{figure}[ht]

\centering
\footnotesize

\includegraphics{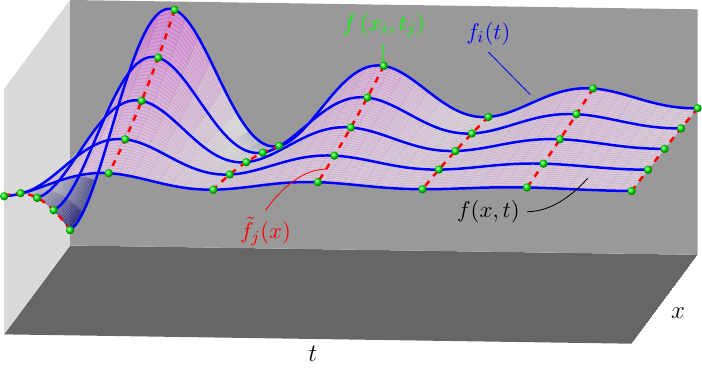}
\caption{Visualisation of a multi-output SML problem}
\label{fig:multi_out}
\end{figure}

A simpler solution of such problems is depicted in Figure \ref{fig:multi_out} on a one dimensional example for visualisation purposes. The continuous time response field $f(x,t)$ is discretized at each of the $t_j \; \forall \; j \in [0,j_{\mathrm{max}}]$ time steps, which is the general state of the data in such problems. Furthermore, the time series $f_i(t) = f(x=x_i,t)$ is measured at all of the sampling points $x_i$, analogous to a single-output problem. Hence, the response data $\mathbf{Y}^0$ is not a vector any more, consisting of a scalar value for each sample, but a matrix, where each row corresponds to a point $x_i$ in the variable space and each column corresponds to a discretized time step $t_j$. Therefore, for the approximation of such a time data field $f(x,t)$, one could train a surrogate model $\tilde{f}_j(x) \approx f(x,t=t_j)$ for each of the time steps $t_j$ and combine the scalar solutions to a vector, corresponding to the response signal. The number of input parameters of the time response field decides the dimensionality of such a problem.

This strategy can become exhausting, if the number of time steps are high. Matrix decomposition methods such as principal component analysis (PCA) can be used to reduce the number of responses by losing some of the variability of the original function. PCA is an orthogonal linear transformation that transforms the data to a new coordinate system without or with limited loss. PCA is also known as the Karhunen-Loeve transformation or the proper orthogonal decomposition depending on the field of application. The dimensions of the new coordinate system are collinear to the corresponding direction in the data space with descending amount of maximum total variance \cite{Pearson1901}. 

By using only a limited number of transformations and directions, the number of dimensions is reduced, which produces some unexplained variance. The unexplained variance can be estimated and used to decide for the right number of components to use for the transformation, which correspond to the number of the responses in the reduced set. In this research the implementation in \cite{SKLEARN} is used for the principal component analysis.

Furthermore, instead of using multiple independent networks, a single neural network is used in DGCN for the estimation of the tuning parameters. Thus, the network does not only learn the dependency between the input parameters and the response at each discretized time step, but also the dependency between each of the time steps implicitly by minimizing the error of all steps simultaneously during the training.

Finally, the small duration of the computation allows greedy optimization algorithms to be employed. In this research, particle swarm optimization \cite{Eberhart1995} (PSO) is used as implemented in \cite{inspyred}, which is a generic nature inspired stochastic optimization algorithm that does not require any gradient information and exhibits a global search behaviour. Application-specific adjustments are not used to increase the efficiency. Instead, the number of candidates in a population as well as the number of total generations are increased for a greedier search of the optimization space. Thus, this algorithm is referred to as a black box optimization method.

\section{Small-Scale Laboratory Experiment}
\label{sec3}
\subsection{The General Setup}
\label{subsec31}
\begin{figure}[ht] 
\captionsetup[subfloat]{captionskip=0.3cm}
\subfloat[Schmematic setup \label{fig:experiment1}]{
\includegraphics[width=0.5\textwidth]{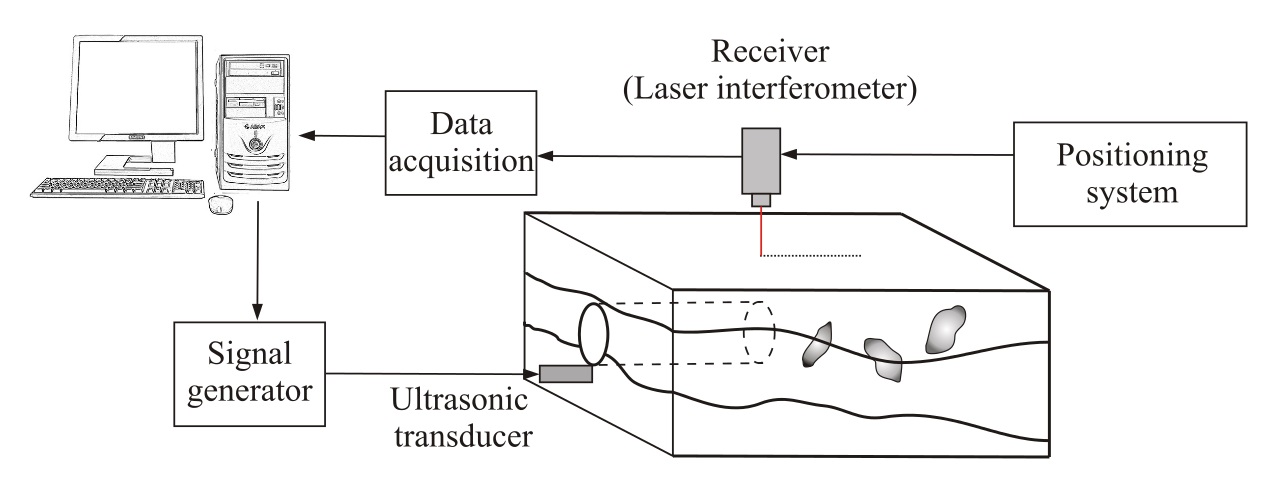}
\vspace*{1em}}\hfill
\subfloat[Photo of the setup \label{fig:experiment2}]{
\includegraphics[width=0.4\textwidth]{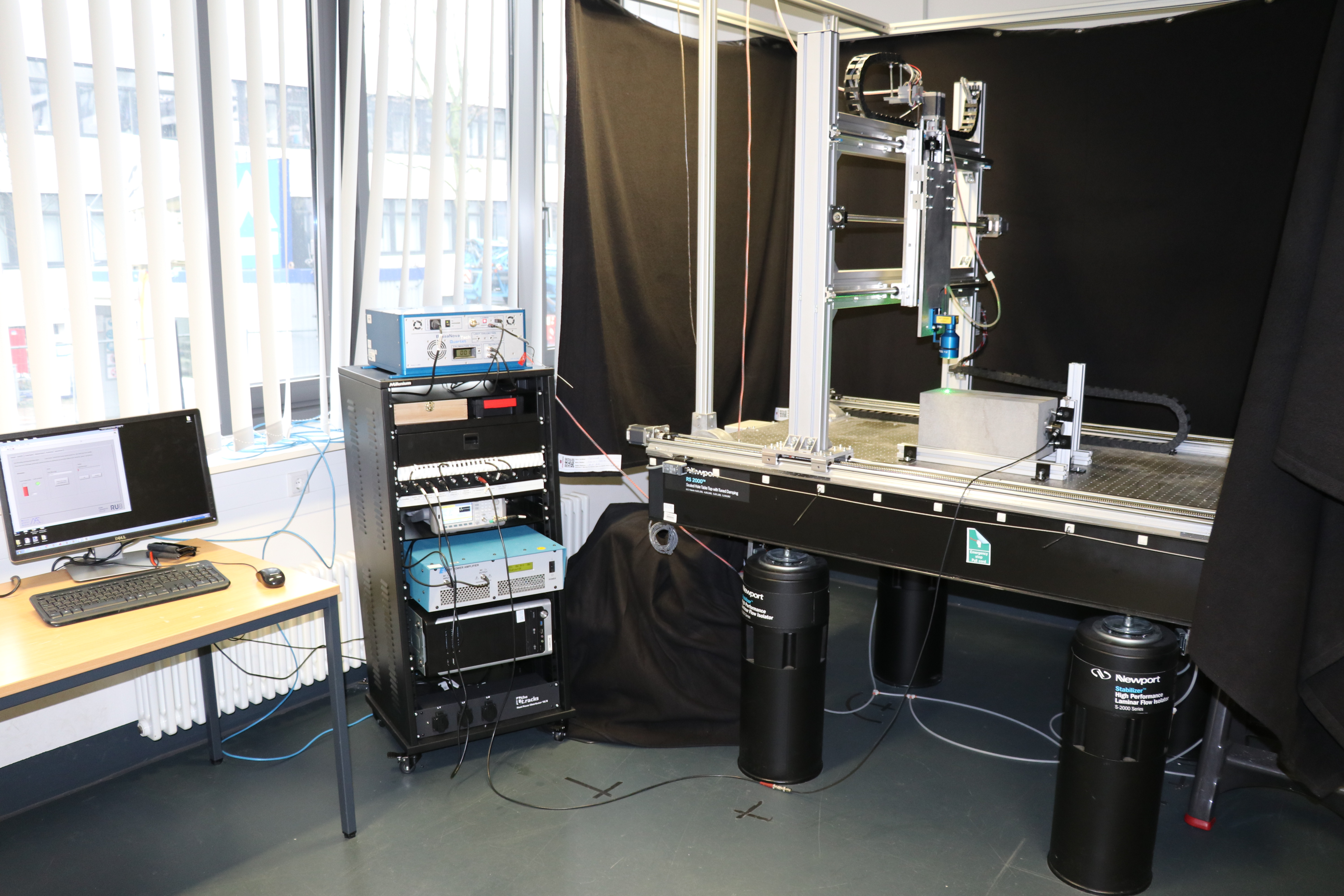}
\vspace*{1em}}
\caption{Small-scale laboratoy experiment}\label{fig:experiment}
\end{figure}

For the purpose of validating the developed inversion methods with experimental data, a small-scale laboratory experiment is built. 
Preparatory to the construction, it is crucial to select individual components compatible to each other. Figure \ref{fig:experiment1} shows the schematic setup, Figure \ref{fig:experiment2} shows a photo of the experimental setup mounted on an optical table for the purpose of vibrational isolation from the environment. The overall system works in the ultrasonic range with frequencies between about $50$ kHz and $2$ MHz.  Ultrasonic transducers trigger the excitation of elastic waves into the investigated medium. For waveform acquisition, the laser interferometer BNT-QUARTET-$500$ linear by Bossa Nova Technologies is used as it enables a contact-free recording of the seismic waves. The navigation of the laser is performed by a positioning system that is controlled by a self-developed software. The software allows to make single-point measurements on the specimen as well as automated measurements along an axis. A Newport optical table serves for holding the experimental setup on a layer of compressed air. This increases the signal-to-noise ratio by reducing influences from the environment.

\subsection{The Experiment}
\label{subsec32}

\begin{figure}[ht]
\subfloat[Disturbed specimen with a drilling of 16 mm diameter. Measurements are taken along the $145$ mm-long green line in $5$ mm-distances. \label{fig:specimen1}]{
\includegraphics[width=0.4\textwidth]{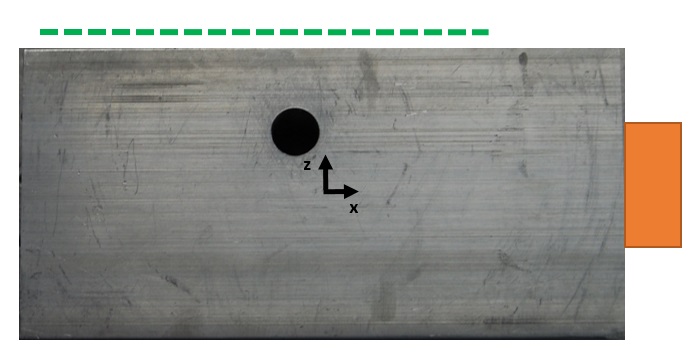}}\hfill%
\subfloat[Intact specimen. A single-shot measurement is taken at the opposite side of the transducer (green cross).\label{fig:specimen2}]{
\includegraphics[width=0.4\textwidth]{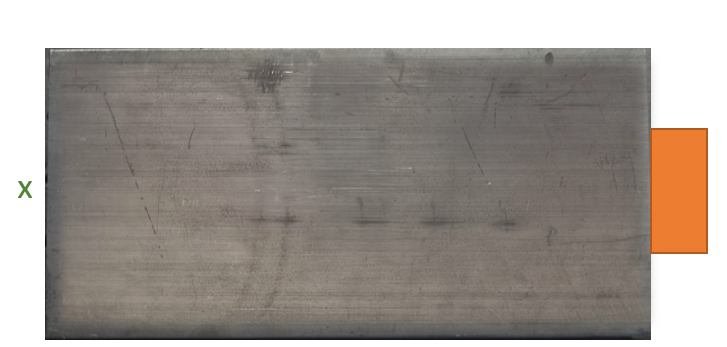}
}
\caption{Investigated specimen with ultrasonic transducers in orange}\label{fig:specimen}
\end{figure}

The aim of the inversion scenario is to localize the coordinates of a borehole in an aluminium specimen. For the investigations, two aluminium blocks with the dimensions of $100\times103\times200.4$ mm are used. Into the first specimen, a hole with a diameter of $16$ mm is drilled. The drilled hole is aimed to be placed at $[-10,20]$ mm with respect to the coordinate system originating in the center of the specimen (Figure \ref{fig:specimen1}), drilling perpendicular to the $xz$-plane. The piezoelectric transducer for the excitation of the ultrasonic wave propagation is placed in the middle of one of the smaller surfaces. The data is aquired along the top surface of the specimen. In order to achieve good measurement results, it is crucial not to measure points which are close to the transducer surface. This is because the signal directivities of ultrasonic transducers are not constant as Bretaudeau et al. show in \cite{Bretaudeau_etal2011}. As the discrepancy between measurements and simulations gets worse with a decreasing direction angle, the receivers are chosen in a way that the maximum angle between the transducer-receiver line and the orthonormal line of the transducer is about $45$ degrees. Consequently, the software is set up so that the laser interferometer acquires data along a $145$ mm long line in $5$ mm-distances along the middle of the top surface, resulting in $30$ points of measurement. 
The second specimen remains intact. For the purpose of later model building, a single-point measurement is taken on the direct opposite of the transducer surface (see Figure \ref{fig:specimen2}). In order to increase the signal-to-noise ratio, all measurements are conducted $500$ times and get stacked. For the wave excitation, a Ricker signal with a central frequency of $300$ kHz is taken as it has a broad frequency spectrum. The ultrasonic transducer used for the investigations is `S 24 HB 0.2-0.6' by the enterprise Karl Deutsch with a flat contact area.

\section{Numerical Model of Setup}
\label{sec4}

In order to achieve a successful inversion, it is crucial to implement a well-fitting numerical model. The geometrical model building and the meshing is realized with Trelis $15.2$ \cite{AmericanFork}; SPECFEM3D \cite{Komatitsch2012} serves as the solver of the wave equations. The latter is published under the GPL $2$ license and uses the continuous Galerkin spectral-element method to simulate forward seismic wave propagation. 

\begin{figure}[ht]
\centering
\includegraphics[width=1.0\textwidth]{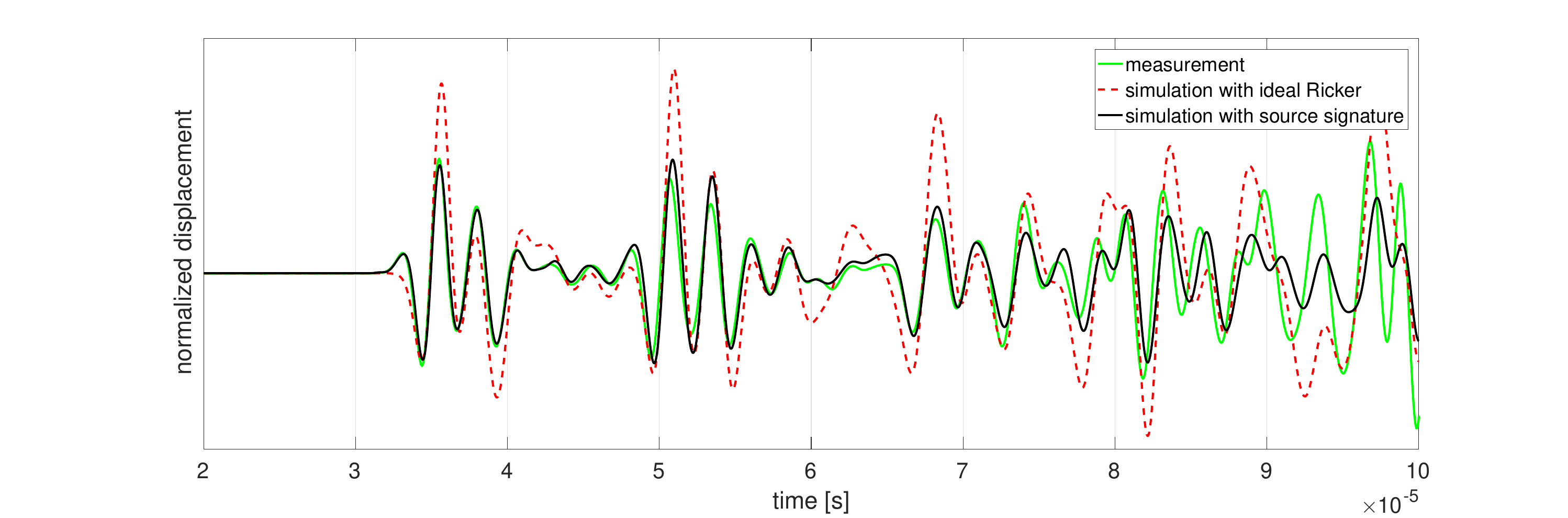}
\centering
\caption{Comparison of measured and simulated waveforms}
\label{fig:seismograms}
\end{figure}

In a first step, a parameter identification is achieved. Therefore, an investigation corresponding to the model of the intact specimen (Figure \ref{fig:specimen2}) is conducted with the aim to maximize the agreement of the waveforms of the simulation and the measurements. The model dimensions of this specimen are measured and based on that, the geometrical model is constructed, i.e.~a brick of dimension $100\times103\times200.4$ mm. All surfaces are defined as free surfaces. The material is considered to be elastic in order to save computational power. Thus, the material parameters are confined to the density and the wave velocities. The density is determined to be $2775~\mathrm{\frac{kg}{m^{3}}}$. Travel time measurements with different transducers are performed to determine the compressional wave velocity. The shear wave velocity is obtained from relations between the compressional wave velocity and an assumed value of $0.34$ for Poisson's ratio. These values are optimized by a sampling investigation, comparing measurements and simulations. The investigations show that values of $6340~\mathrm{\frac{m}{s}}$ for compressional wave velocity and $3110~\mathrm{\frac{m}{s}}$ for shear wave velocity are well-fitting, which can be seen in Figure \ref{fig:seismograms}. To remain stable, the model shall be discretised in a way that there are at least 5 sampling points per wavelength. As SpecFEM3D uses a standard preference of 5 Gauss-Lobatto-Legendre points per element edge, the mesh size is chosen so that $l_{max}<\frac{v_{s}}{f_{max}}$, where $l_{max}$ is the largest edge of an element in the model, $v_{s}$ is the shear wave velocity and $f_{max}$ is the maximum occurring frequency. The maximum frequency is set to $2.5 \times f_{c}$, where $f_{c}$ is the central frequency of the Ricker signal. Running the database generator of SpecFEM3D, a maximum suggested time step is proposed by the software which is used for the simulations. To consider the effect of the flat transducer, sources are applied on every node of the model with which the transducer would be in contact. 

A first simulation with an ideal Ricker signal of $300$ kHz is made. Figure \ref{fig:seismograms} shows the simulation result in red dashed and the corresponding single-point measurement in green.

\begin{figure}[ht]
\center 
\includegraphics[width=1.0\textwidth]{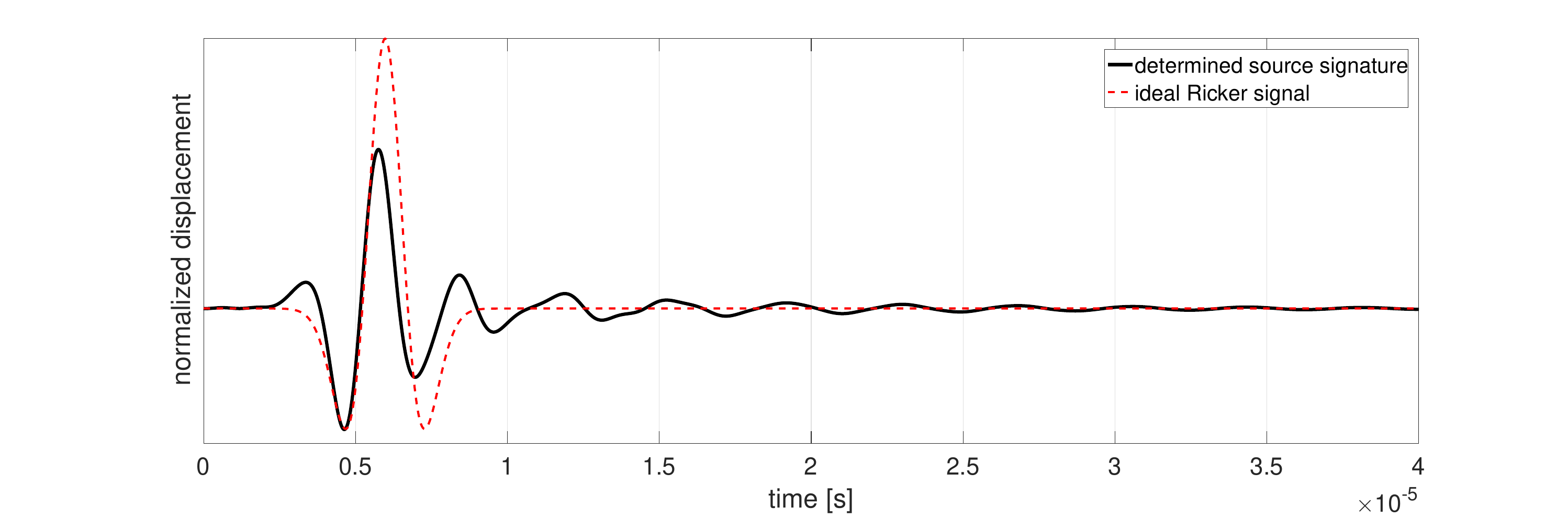}
\centering
\caption{ideal Ricker signal and determined source signature}
\label{fig:sourcesignature}
\end{figure} 

It is conspicuous that the waveforms are very similar which shows that the material parameters are well estimated. However, to get an optimal estimation, it is inevitable to determine the source signature of the system.~The source signal differs from the ideal Ricker signal as it depends on the transducer's coupling with the structure and its technical properties. Assuming that there is a perfect agreement between the simulation model and the real model, the following applies:
\begin{equation} 
Q(\omega)=\frac{D(\omega)}{S(\omega)} R(\omega), 
\end{equation}
where $Q(\omega)$ is the Fourier transform of the sought source function, $D(\omega)$ the Fourier transform of the measurement data, $S(\omega)$ the Fourier transform of the obtained simulation results with an ideal Ricker signal and $R(\omega)$ the Fourier transform of the ideal Ricker wavelet. Possessing the measurement data and the simulation data for the ideal case, one can derive the source signature in time domain by performing an inverse Fourier transform. Figure \ref{fig:sourcesignature} shows the derived source signature compared to the ideal source signature. Figure \ref{fig:seismograms} shows the simulation results generated with the determined source wavelet.~It is obvious that the determination of the source signature strongly improves the results as the agreement is almost perfect. Thus, well-fitting model parameters and the source signature are derived and used for the inversion problem.

\section{Anomaly Detection}
\label{sec5}

The aim of this section is the detection of the drilling hole coordinates of the aluminium specimen (Figure \ref{fig:specimen}) with the three methods presented in Section \ref{sec2}. In each case, the optimum is sought by minimizing the misfit function (Eq.~\ref{eqn:misfit}) which represents the difference between the measurement data (see Section \ref{sec3}) and the simulation data (see Section \ref{sec4}) for the current coordinate of the drilling hole. A simulation takes $7-8$ minutes on average on a 26-core computer with 2.4 GHz each and 96 GB RAM. Before calculating the least square difference, the simulation data is resampled to the measurement sampling rate. The amplitude of the simulation data is then aligned to the amplitude of the experimental data by calculating the factor which leads to a minimum gap. The duration of the investigated signal is chosen to be $8.8 \times 10^{-5}$ s as Figure \ref{fig:seismograms} shows a good fit of the waveforms up to that time. The calculation of the misfit is finally done with this processed signal. 

\begin{figure}[ht]
\setlength\fheight{0.4\textwidth} 
\setlength\fwidth{\textwidth}

\includegraphics{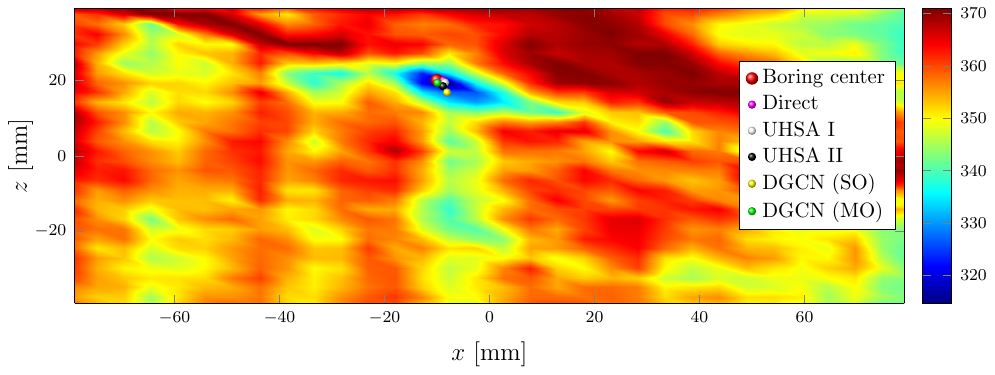}

\caption{Linear interpolation of the landscape using 1024 points}\label{fig:Landscape}
\end{figure}

Figure \ref{fig:Landscape} shows the misfit landscape of the problem which is determined by a linear interpolation of 1024 points. The abscissa corresponds to the $x$-axis in Figure \ref{fig:specimen1} and the ordinate to the $z$-axis. The legend entry boring center describes the estimated position of the borehole. Note that this position is not the true optimum as it is only the position which was aimed when drilling the hole. Direct corresponds to the strategy of coupling PSO with the simulation model.~UHSA I and UHSA II depict the results of the UHSA algorithm with different settings.~DGCN (SO) and DGCN (MO) show the results of DGCN using the misfit as a single-output and the whole signal as multiple outputs respectively. It can be observed that the misfit response function has only a single global minimum in a local neighbourhood. The individual results are further discussed in the following.

\subsection{Direct Coupling of the Simulation Model with the Black Box Optimization}
For this approach, the computationally expensive numerical model is directly coupled with PSO. The misfit function is minimized within the geometric bounds. Because of the amount of the computational effort and the limited resources, the number of generations is set to $12$ with a population size of $50$, which is relatively small for this algorithm. It took approximately $70$ hours of computation time on the high performance computing cluster described above.

During the preliminary tests, the tendency of the gradient based algorithms to converge to wrong local minima is observed. Although gradient based methods can be more efficient, PSO is chosen because of its global search behaviour as mentioned at the end of Section \ref{sec:MOSML}. After $600$ calculations, a relatively good approximation of the the global optimum is found, as it can be see in Figure \ref{fig:Landscape} and Table \ref{tab:AllResults}. The results of this method are included for the comparison of its efficiency. It needed more than twice as much calculations, compared to the first configuration of UHSA, which required the most number of calculations among the rest of the proposed strategies. 

\subsection{Results of Unscented Hybrid Simulated Annealing}
\label{subsec51}

In order to execute the UHSA optimization, the setting of an input configuration is required. SA is tuned with the parameters ${T}_0$ and $\alpha$ (see Section \ref{subsec213}); the UKF needs $\bm{P}_0^m$, $\bm{R}$, $\bm{Q}$, $\bm{s}_{min}$, $N_{ukf}$ for initialization (see Section \ref{subsec212}). Additionally, the bounds of the input parameters are set to $[-80, 80]$ mm in $x$-direction and to $[-40, 40]$ mm in $z$-direction according to the coordinate system specified in Figure \ref{fig:specimen}. These bounds mark the possible area for the coordinates of the hole center in the inversion scenario. If these bounds are exceeded, the algorithm jumps to the next cycle.  In order to adjust a covariance matrix $\bm{R}$ and a vector for the best expected result $\bm{s}_{min}$, a vector $\bm{s}_{u}$ is determined. This vector contains the misfit at each receiver between the experimental data used for inversion and the simulation data of an undisturbed specimen. $\bm{R}$ and $\bm{s}_{min}$ can then be tuned with optional parameters $a,b$ setting $\bm{s}_{min}=a \cdot \bm{s}_{u}$ and $\bm{R}=b  \cdot\bm{s}_{u}\bm{I}_r$, where $\bm{I}_r \in \mathbb{R}^{r\times r}$ is an identity matrix of dimension equal to the number of receivers; in this case $r=30$. With this operation, the parameters can be set rather intuitively as the best misfit can be expected to be smaller than the misfit of an undisturbed specimen and thus $a,b \in{[0,1]}$. Furthermore, the quality of the measurement at each receiver point is considered. Note that the parameters can also be tuned by setting $\bm{s}_{min}$ to a zero vector and the measurement covariance $\bm{R}$ rather high, but doing the former is more intuitive.

\begin{table}[!htb]
\centering
\begin{tabular}{l|c|c}
	Parameter							& Configuration 1	& Configuration 2 \\ [0.5ex]
	\hline\hline 
& & \\ [-2ex]
$\bm{P}_0^m$ 	& $\textrm{diag}({4}^2, {4}^2)$ 			& $\textrm{diag}({2}^2, {2}^2)$ 	\\
	$\bm{Q}$  					& $0.1 \cdot \bm{P}_0^m$			& $0.1 \cdot \bm{P}_0^m$	\\
	$\bm{s}_{min}$  					& $0.8 \cdot \bm{s}_{u}$			& $0.7 \cdot \bm{s}_{u}$	\\
	$\bm{R}$	&	 $0.2  \cdot\bm{s}_{u}$   &  $0.1  \cdot\bm{s}_{u}$\\
	$N_{ukf}$  					& 	$4$		& 	$2$ \\
	$T_{0}$							  					& 	$9.49$		& 	$1.44$  \\
	$\alpha$														  					& 	$0.95$		& $0.9$	\\ [0.5ex]
\end{tabular}
\caption{Input configurations for UHSA}
\label{tab:configs}
\end{table}

\begin{figure}[!ht]
\centering
\setlength\fheight{0.4\textwidth} 
\setlength\fwidth{0.9\textwidth}
\subfloat[Configuration 1 \label{fig:contours1}]{

\includegraphics{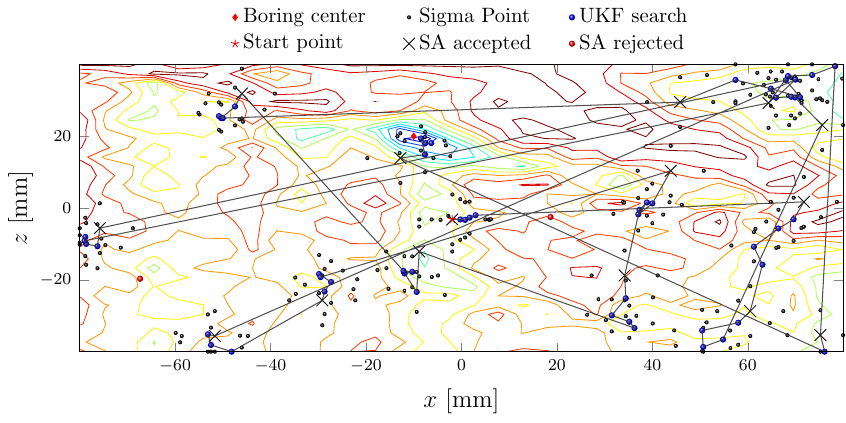}
} \\
\setlength\fheight{0.4\textwidth} 
\subfloat[Configuration 2 \label{fig:contours2}]{
\vspace{0.5cm}

\includegraphics{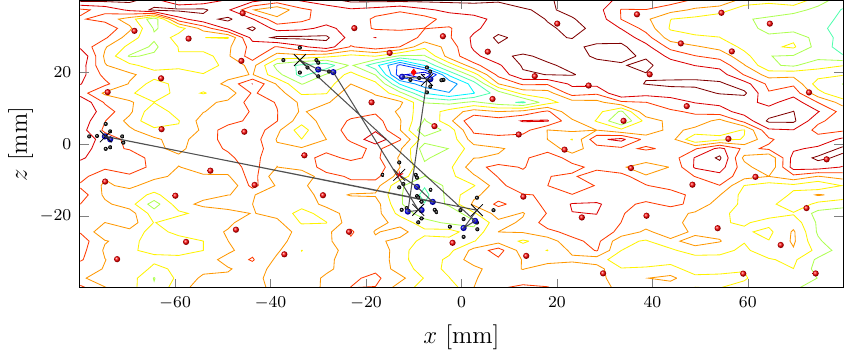}
}
\caption{Course of the UHSA in the optimization space, plotted onto the predicted misfit landscape}\label{fig:contours}
\end{figure}

The UHSA algorithm is set for two examples of input configurations which are shown in Table \ref{tab:configs}. Configuration $1$ is set with the aim to find the global minimum accurately and in an early stage of annealing cycles. Therefore, a high annealing temperature is set which corresponds to an average acceptance probability of $0.9$ in the beginning. The sigma points are widely spread by large values in $\bm{P}_0^m$ (covariance refers to a random variable in mm-unit here), meaning that an accepted value of SA does not have to lie in the direct global minimum region for a successful optimization. The setting of a relatively high number of Kalman runs secures that the minimum region is exploited accurately. $\bm{s}_{min}$ and  $\bm{R}$ are chosen intuitively -- their adjustment influences the distance of a jump to the next run in a Kalman cycle. In order to avoid sample points to be located too close to each other, every proposed configuration by SA is set to be $10$ mm distant from each proposed configuration beforehand. \\
Configuration $2$ is set with the aim to find the optimum with less computational effort and therefore with a smaller focus on accuracy. A strict annealing schedule is set that corresponds to an average acceptance probability of $0.5$ in the beginning. Thus, computational power is saved as not many configurations are accepted for local minimization. The consequence is that the proposed parameter configuration has to be close to the global minimum region as the strict annealing cycle would hinder it from being accepted instead. Thus, the values in the covariance matrix $\bm{P}_0^m$ and the number of Kalman runs can be set low. $\bm{s}_{min}$ and $\bm{R}$ are chosen slightly smaller than in the first example to avoid jumps getting too small. The minimum distance between the proposed configurations is set to $5$ mm. In both examples, the algorithm stops if the value of the objective function gets lower than a threshold.

Figure \ref{fig:contours} shows the courses of the algorithms on the parameter landscape, where the abscissa corresponds to the $x$-axis in Figure \ref{fig:specimen} and the ordinate to the $z$-axis. Figure \ref{fig:contours1} shows the course for configuration $1$. It can be observed that nearly every proposed annealing cycle has been accepted. The exploration of local minima is quite intense, meaning that a high number of forward simulations is needed. In cycle $17$, the accepted point is close enough to the global minimum for the algorithm to exploit it. The minimum is found after $293$ forward simulations at $[-8.46, 19.37]$ mm, deviating a distance of only $1.66$ mm from the approximated position of the borehole of $[-10,20]$ mm. \\
The course of UHSA for input configuration $2$, which is illustrated in Figure \ref{fig:contours2}, shows less affinity for acceptance. In the overall course, 6 proposed samples are accepted. Only $122$ forward simulations are performed with the best parameter configuration of $[-8.84, 18.39]$ mm, deviating a distance of $1.98$ mm from the approximated position of the borehole. The results are summarized in Table \ref{tab:AllResults}.\\
It can be observed that the second configuration reduces the computational effort massively, forfeiting an acceptable amount of precision. Thus, for a low-parameter space as in this example, it seems that a strict annealing cycle and weak local exploration are the better choice if the computational effort is an issue. For higher-parameter spaces like in tunneling problems, it can be assumed that higher values for the annealing parameters have to be chosen in order to prevent the global minimum from being missed in an early stage.

\subsection{Black Box Optimization using DGCN}

\begin{figure}[ht]
\setlength\fheight{0.4\textwidth} 
\setlength\fwidth{0.54\textwidth}
\subfloat[Multi-Output \label{fig:DGCNMO}]{

\includegraphics{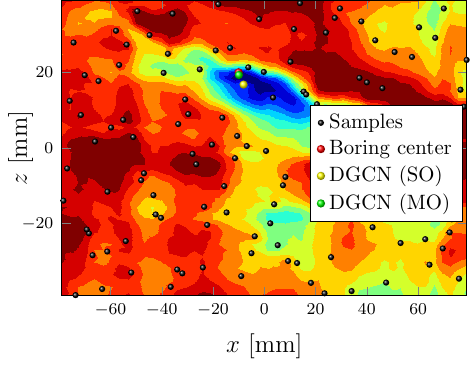}
}\hfill 
\setlength\fheight{0.4\textwidth} 
\subfloat[Single-Output \label{fig:DGCNSO}]{
\includegraphics{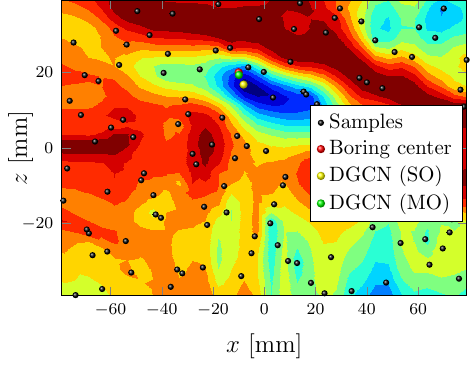}
}
\caption{Predicted landscape of the misfit with single and multi-output DGCN models using 128 samples}\label{fig:DGCNMOSO}
\end{figure}

The general strategy of employing machine learning is conducting a number of forward simulations to generate data and using the data to acquire a surrogate model. In this example, $128$ samples are generated using an optimized Latin hypercube scheme \cite{Joseph2008}. Hence, $128$ signals are calculated and a multi-output DGCN model is trained for each of the $30$ sensors (Figure \ref{fig:DGCNMO}). Furthermore, the misfit is calculated for each sampling point and a single-output DGCN model is trained using the misfit as the response (Figure \ref{fig:DGCNSO}). There are slight differences in the landscape, although both models use the same method and the same inputs, aside from the response values.

Since only stationary sampling is used, all of the samples could be computed in parallel, reducing the computational time to approximately $7$ minutes, as far as enough computational resources are available, which means in this case $128 \times 26$ cpu cores for the calculation. As this is not the case during this research, the simulations are calculated sequentially and the total computational time is approximately $15$ hours. The efficiency of this method can be further increased by using adaptive sampling schemes such as Bayesian optimization as proposed in \cite{Mockus1994}.

Finally, PSO with $200$ generations with $200$ individuals is deployed on the resulting surrogate models and the misfit is minimized within the aforementioned bounds. In both cases, PSO converges before the $50$-th generation. The convergence is defined by the stagnation of the objective value over $15$ generations. The results are depicted in Figure \ref{fig:Landscape} and given in Table \ref{tab:AllResults}. Error margins in both results are acceptably small for this application but the results of multi-output DGCN are slightly better than the single-output DGCN. This may look surprising, since the single-output problem is considered to be easier from a machine learning perspective, because there is only a single response to approximate. However, since the misfit is an integral response of all errors, small perturbations in the signals may be interpreted as noise and smoothed away because of more dominant modes of the signal. If such perturbations are relevant for the solution of the inverse problem, the difference in the quality of the results are explainable. Nevertheless, the multi-output model can replace the simulation model as it is and be used with any future measurements, where as a new single-output model must be trained for each new measurement, which can be unaffordable for continuous measurement applications. Therefore using multi-output SML may be more advantageous in time dependent applications. 

\begin{table}%
\centering
\begin{tabular}{l|c|c|c}
Strategy & Coordinates [mm] & Misfit & No. of calculations\\ [0.5ex]
\hline \hline
& & \\ [-2ex]
PSO (Direct) & $[-8.79,19.96]$ &  $296.91$ & $600$ \\
DGCN (SO) & $[-8.08,16.75]$ &  $321.54$ & $128$\\
DGCN (MO) & $[-9.92,19.29]$ &  $306.27$ & $128$\\
UHSA I & $[-8.46,19.37]$ &  $297.02$ & $293$ \\
UHSA II & $[-8.84,18.39]$ &  $304.74$ & $122$\\ [0.5ex]
\end{tabular}
\caption{Results of the strategies} \label{tab:AllResults}
\end{table}

\section{Conclusion}
\label{sec6}

The novel unscented hybrid simulated annealing algorithm is tested against the black box optimization methods. It is shown that UHSA finds a solution that is approximately as good as the direct coupling of the simulation model to the black box optimization algorithm PSO. Simultaneously, UHSA is much more effective, needing only 293 or 122 simulations instead of 600 simulations required by the direct coupling. The precision can be further improved by setting the UKF exploration larger or by doing an additive localization run after the UHSA optimization. The tuning of the parameters for the UHSA makes it possible to adjust the inversion to the requirements of the user, finding a compromise between the computational effort and the precision. However, due to the non-deterministic nature of the algorithm, the inversion time is always partly coincidential. Thus, the optimum can be found even faster, but also slower. \\
Furthermore, machine learning is used to increase the efficiency of the black box optimization strategy. Using single and multi-output machine learning, the efficiency of the black box optimization increases, requiring only $128$ forward calculations instead of the original $600$ required by the direct coupling but some deviation is observed in the results, which is sufficiently small for this application. Furthermore, it is observed that better results are achieved by using a multi-output formulation of the machine learning problem as opposed to the more conventional single-output formulation. UHSA can be as efficient as deploying machine learning with less deviation from the optimum, which makes it the best strategy for this problem.\\ Nevertheless, there are some advantages of using DGCN compared to UHSA. Since stationary sampling is used, all calculations can be done in parallel, as long as sufficient amount of computational resources are available. For UHSA, the number of possible parallel simulations is limited to the number of sigma points, which is $5$ in this example due to the two-dimensionality of the inversion problem. The maximum number of parallel calculations for the direct strategy is limited to the number of individuals in a generation of the PSO. Furthermore, using multi-output DGCN, the surrogate model is independent of the measurement, hence it can replace the simulation model completely as opposed to the single-output strategy, where the misfit heavily depends on the measurement. This can be useful for on-site applications, where continuous measurement is conducted. Consequently, replacing the numerical model in UHSA with multi-output DGCN is another possible strategy, that is left out of this research as this would yield little to no gain when the number of measurement sets is small. The combination would only be meaningful in continuous measurement settings. \\
Finally, a homogeneous aluminium brick is used throughout the investigation, which is much easier to model compared to the soil material found in the target applications, as discussed at the end of the introduction. However, as this is a modeling problem, the proposed strategies are expected to work in such a setting, as long as the model captures the real properties of the investigated material. The possible decrease of efficiency caused by a higher number of dimensions and a higher level of nonlinearity in such cases should be further investigated in future work.

\section*{Acknowledgements}

The authors gratefully acknowledge the funding by the German Research Foundation (DFG) within the Collaborative Research Center SFB 837 "Interaction modeling in mechanized tunneling/ Sub-project A2: Development of effective concepts for tunnel reconnaissance using acoustic methods."

\appendix

%



\bibliographystyle{elsarticle-num}
\bibliography{LiteratureBogoclu,sample}

\begin{thebibliography}{10}
\expandafter\ifx\csname url\endcsname\relax
  \def\url#1{\texttt{#1}}\fi
\expandafter\ifx\csname urlprefix\endcsname\relax\def\urlprefix{URL }\fi
\expandafter\ifx\csname href\endcsname\relax
  \def\href#1#2{#2} \def\path#1{#1}\fi

\bibitem{Schmitt2004}
J.~Schmitt, J.~Gattermann, J.~Stahlmann, Hohlraumerkundung im {T}unnelbau,
  Messen in der Geotechnik (2004) 173--200.

\bibitem{Sattel1992}
G.~Sattel, P.~Frey, R.~Amberg, Prediction ahead of the tunnel face by seismic
  methods-pilot project in {C}entovalli tunnel, {L}ocarno, {S}witzerland, First
  Break 10~(1) (1992) 19--25.

\bibitem{Kneib_etal2000}
G.~Kneib, A.~Kassel, K.~Lorenz, Automatic seismic prediction ahead of the
  tunnel boring machine, First Break 18~(7) (2000) 295--302.

\bibitem{Otto_etal2002}
R.~Otto, E.~Button, H.~Bretterebner, P.~Schwab, The application of trt (true
  reflection tomography) at the {U}nterwald tunnel, Felsbau 20~(2) (2002)
  51--56.

\bibitem{Borm_etal2003}
G.~Borm, R.~Giese, C.~Klose, S.~Mielitz, P.~Otto, T.~Bohlen, {ISIS} integrated
  seismic imaging system for the geological prediction ahead of underground
  construction, in: EAGE the 65th Conference and Exhibition. Norway:[sn], 2003.

\bibitem{Tarantola1984}
A.~Tarantola, Inversion of seismic reflection data in the acoustic
  approximation, Geophysics 49~(8) (1984) 1259--1266.

\bibitem{Musayev2014}
K.~Musayev, K.~Hackl, M.~Baitsch, Identification of the velocity field of 2d
  and 3d tunnel models with frequency domain full waveform inversion, PAMM
  14~(1) (2014) 781--782.

\bibitem{Bretaudeau_etal2013}
F.~Bretaudeau, R.~Brossier, D.~Leparoux, O.~Abraham, J.~Virieux, 2d elastic
  full-waveform imaging of the near-surface: application to synthetic and
  physical modelling data sets, Near Surface Geophysics 11~(3) (2013) 307--316.

\bibitem{Bretaudeau_etal2014}
F.~Bretaudeau, C.~G\'elis, D.~Leparoux, R.~Brossier, J.~Cabrera, P.~Côte,
  \href{http://library.seg.org/doi/abs/10.1190/geo2013-0082.1}{High-resolution
  quantitative seismic imaging of a strike-slip fault with small vertical
  offset in clay rocks from underground galleries: Experimental platform of
  {T}ournemire, {F}rance}, GEOPHYSICS 79~(1) (2014) B1--B18.
\newblock \href
  {http://arxiv.org/abs/http://library.seg.org/doi/pdf/10.1190/geo2013-0082.1}
  {\path{arXiv:http://library.seg.org/doi/pdf/10.1190/geo2013-0082.1}}, \href
  {http://dx.doi.org/10.1190/geo2013-0082.1}
  {\path{doi:10.1190/geo2013-0082.1}}.
\newline\urlprefix\url{http://library.seg.org/doi/abs/10.1190/geo2013-0082.1}

\bibitem{du2017microstructure}
H.~Du, K.~Carpenter, D.~Hui, M.~Radonjic, Microstructure and micromechanics of
  shale rocks: Case study of marcellus shale, Facta Universitatis, Series:
  Mechanical Engineering 15~(2) (2017) 331--340.

\bibitem{Lamert2018}
A.~Lamert, W.~Friederich, Full waveform inversion for advance exploration of
  ground properties in mechanized tunneling, International Journal of Civil
  Engineering (2018) 1--14.

\bibitem{Nguyen2016}
L.~T. Nguyen, T.~Nestorovi{\'c}, Unscented hybrid simulated annealing for fast
  inversion of tunnel seismic waves, Computer Methods in Applied Mechanics and
  Engineering 301 (2016) 281--299.

\bibitem{Julier&Uhlmann2004}
S.~J. Julier, J.~K. Uhlmann, Unscented filtering and nonlinear estimation,
  Proceedings of the IEEE 92~(3) (2004) 401--422.

\bibitem{Kirkpatrick_etal1983}
S.~Kirkpatrick, M.~Vecchi, et~al., Optimization by simmulated annealing,
  science 220~(4598) (1983) 671--680.

\bibitem{Brigham2007}
J.~C. Brigham, W.~Aquino,
  \href{http://www.sciencedirect.com/science/article/pii/S0045782507002265}{Surrogate-model
  accelerated random search algorithm for global optimization with applications
  to inverse material identification}, Computer Methods in Applied Mechanics
  and Engineering 196~(45) (2007) 4561 -- 4576.
\newblock \href {http://dx.doi.org/https://doi.org/10.1016/j.cma.2007.05.013}
  {\path{doi:https://doi.org/10.1016/j.cma.2007.05.013}}.
\newline\urlprefix\url{http://www.sciencedirect.com/science/article/pii/S0045782507002265}

\bibitem{Bilicz2012}
S.~Bilicz, M.~Lambert, S.~Gyimothy, J.~Pavo, Solution of inverse problems in
  nondestructive testing by a kriging-based surrogate model, IEEE Transactions
  on Magnetics 48~(2) (2012) 495--498.
\newblock \href {http://dx.doi.org/10.1109/TMAG.2011.2172196}
  {\path{doi:10.1109/TMAG.2011.2172196}}.

\bibitem{Marzouk2009}
Y.~M. Marzouk, H.~N. Najm,
  \href{http://www.sciencedirect.com/science/article/pii/S0021999108006062}{Dimensionality
  reduction and polynomial chaos acceleration of bayesian inference in inverse
  problems}, Journal of Computational Physics 228~(6) (2009) 1862 -- 1902.
\newblock \href {http://dx.doi.org/https://doi.org/10.1016/j.jcp.2008.11.024}
  {\path{doi:https://doi.org/10.1016/j.jcp.2008.11.024}}.
\newline\urlprefix\url{http://www.sciencedirect.com/science/article/pii/S0021999108006062}

\bibitem{Aster2013}
R.~C. Aster, B.~Borchers, C.~H. Thurber,
  \href{http://www.sciencedirect.com/science/article/pii/B9780123850485000094}{Chapter
  nine - nonlinear regression}, in: R.~C. Aster, B.~Borchers, C.~H. Thurber
  (Eds.), Parameter Estimation and Inverse Problems (Second Edition), second
  edition Edition, Academic Press, Boston, 2013, pp. 217 -- 238.
\newblock \href
  {http://dx.doi.org/https://doi.org/10.1016/B978-0-12-385048-5.00009-4}
  {\path{doi:https://doi.org/10.1016/B978-0-12-385048-5.00009-4}}.
\newline\urlprefix\url{http://www.sciencedirect.com/science/article/pii/B9780123850485000094}

\bibitem{Feder1988}
M.~Feder, E.~Weinstein, Parameter estimation of superimposed signals using the
  em algorithm, IEEE Transactions on Acoustics, Speech, and Signal Processing
  36~(4) (1988) 477--489.
\newblock \href {http://dx.doi.org/10.1109/29.1552}
  {\path{doi:10.1109/29.1552}}.

\bibitem{Santamarina2005}
J.~Santamarina, F.~Dante, Discrete Signals and Inverse Problems: An
  Introduction for Engineers and Scientists, John Wiley \& Sons, 2005.
\newblock \href {http://dx.doi.org/10.1002/0470021896}
  {\path{doi:10.1002/0470021896}}.

\bibitem{Simon2006}
D.~Simon, Optimal state estimation: Kalman, H infinity, and nonlinear
  approaches, John Wiley \& Sons, 2006.

\bibitem{VanDerMerwe2004}
R.~Van Der~Merwe, Sigma-point kalman filters for probabilistic inference in
  dynamic state-space models, Ph.D. thesis, Oregon Health \& Science University
  (2004).

\bibitem{Metropolis_etal1953}
N.~Metropolis, A.~W. Rosenbluth, M.~N. Rosenbluth, A.~H. Teller, E.~Teller,
  Equation of state calculations by fast computing machines, The journal of
  chemical physics 21~(6) (1953) 1087--1092.

\bibitem{Balling1991}
R.~J. Balling, Optimal steel frame design by simulated annealing, Journal of
  Structural Engineering 117~(6) (1991) 1780--1795.

\bibitem{Blum&Roli2003}
C.~Blum, A.~Roli, Metaheuristics in combinatorial optimization: Overview and
  conceptual comparison, ACM Computing Surveys (CSUR) 35~(3) (2003) 268--308.

\bibitem{Mitchell1997}
T.~M. Mitchell, Machine Learning, McGraw - Hill Science\/Engineering\/Math,
  1997.

\bibitem{Bogoclu2016}
C.~Bogoclu, D.~Roos, A benchmark of contemporary metamodeling algorithms, in:
  VII European Congress on Computational Methods in Applied Sciences and
  Engineering (ECCOMAS) Congress, Vol.~2, 2016, pp. 3344--3360.

\bibitem{Rasmussen2006}
C.~E. Rasmussen, C.~K.~I. Williams, Gaussian Processes for Machine Learning,
  MIT Press, 2006.

\bibitem{Hastie2009}
T.~Hastie, R.~Tibshirani, J.~Friedman, The Elements of Statistical Learning,
  Springer, 2009.

\bibitem{Cremanns2017}
K.~Cremanns., D.~Roos, \href{https://arxiv.org/pdf/1710.06202.pdf}{Deep
  {G}aussian covariance network}, arXiv:1710.06202v2 [cs.LG] (2017).
\newline\urlprefix\url{https://arxiv.org/pdf/1710.06202.pdf}

\bibitem{Pearson1901}
K.~Pearson, On lines and planes of closest fit to systems of points in space,
  Philosophical Magazine 2~(11) (1901) 559--572.

\bibitem{SKLEARN}
F.~Pedregosa, G.~Varoquaux, A.~Gramfort, V.~Michel, B.~Thirion, O.~Grisel,
  M.~Blondel, P.~Prettenhofer, R.~Weiss, V.~Dubourg, J.~Vanderplas, A.~Passos,
  D.~Cournapeau, M.~Brucher, M.~Perrot, A.~Duchesnay, Scikit-learn: Machine
  learning in python, Journal of Machine Learning Research 12 (2011)
  2825--2830.

\bibitem{Eberhart1995}
R.~C. Eberhart, J.~Kennedy, A new optimizer using particle swarm theory, in:
  MHS'95. Proceedings of the Sixth International Symposium on Micromachine and
  Human Science, IEEE, Nagoya, Japan. IEEE Service Center Piscataway NJ, 1995,
  pp. 39--43.

\bibitem{inspyred}
A.~Garrett, \href{https://pypi.python.org/pypi/inspyred}{inspyred: Bio-inspired
  algorithms in python} (2014).
\newline\urlprefix\url{https://pypi.python.org/pypi/inspyred}

\bibitem{Bretaudeau_etal2011}
F.~Bretaudeau, D.~Leparoux, O.~Durand, O.~Abraham, Small-scale modeling of
  onshore seismic experiment: A tool to validate numerical modeling and seismic
  imaging methods, Geophysics 76~(5) (2011) T101--T112.

\bibitem{AmericanFork}
U.~c. American~Fork, Trelis (version 15.2) [computer software].

\bibitem{Komatitsch2012}
D.~Komatitsch, J.-P. Vilotte, J.~Tromp, J.-P. Ampuero, K.~Bai, P.~Basini,
  C.~Blitz, E.~Bozdag, E.~Casarotti, J.~Charles, M.~Chen, P.~Galvez,
  D.~Goddeke, V.~Hjorleifsdottir, J.~Labarta, N.~Le~Goff, P.~Le~Loher,
  M.~Lefebvre, Q.~Liu, Y.~Luo, A.~Maggi, F.~Magnoni, R.~Martin, R.~Matzen,
  D.~McRitchie, M.~Meschede, P.~Messmer, D.~Michea, S.~Nadh~Somala,
  T.~Nissen-Meyer, D.~Peter, M.~Rietmann, E.~de~Andrade, B.~Savage,
  B.~Schuberth, A.~Sieminski, L.~Strand, C.~Tape, Z.~Xie, H.~Zhu, Specfem3d
  cartesian v2.0.2 [software] (2012).
\newblock \href {http://dx.doi.org/http://doi.org/NoDOI}
  {\path{doi:http://doi.org/NoDOI}}.

\bibitem{Joseph2008}
R.~V. Joseph, Y.~Hung, Orthogonal-maximin latin hypercube designs, Statistica
  Sinica 18 (2008) 171--186.

\bibitem{Mockus1994}
J.~Mockus, Application of bayesian approach to numerical methods of global and
  stochastic optimization, J. Global Optimization 4~(4) (1994) 347--365.

\end{thebibliography}

\end{document}